\documentclass[12pt,fleqn]{article}

\usepackage{amsmath,amssymb,amsfonts,graphicx,epsfig,cite}
\usepackage[footnotesize,width=0.97\textwidth]{caption}
\def\bea{\begin{eqnarray}}
\def\eea{\end{eqnarray}}
\def\be{\begin{equation}}
\def\ee{\end{equation}}
\def\ba{\begin{array}}
\def\ea{\end{array}}
\def\nn{\nonumber}
\def \lsim{\mathrel{\vcenter
{\hbox{$<$}\nointerlineskip\hbox{$\sim$}}}}
\def \gsim{\mathrel{\vcenter
{\hbox{$>$}\nointerlineskip\hbox{$\sim$}}}}

\def\b{&\hspace{-10pt}}

\begin{document}
\pagestyle{plain}

\renewcommand{\theequation}{\arabic{section}.\arabic{equation}}
\setcounter{page}{1}

\begin{titlepage} 
\begin{center}

$\;$

\vskip 0.5 cm

\begin{center}
{\LARGE{ \bf Scalar masses in general N=2 \\[1mm] gauged supergravity theories}}
\end{center}

\vskip 1cm

{\large 
Francesca Catino, Claudio A. Scrucca and Paul Smyth
}

\vskip 0.5cm

{\it Institut de Th\'eorie des Ph\'enom\`enes Physiques, EPFL, \\
\mbox{CH-1015 Lausanne, Switzerland}\\
}

\vskip 1.5cm

\begin{abstract}

We readdress the question of whether any universal upper bound exists
on the square mass $m^2$ of the lightest scalar around a supersymmetry breaking 
vacuum in generic N=2 gauged supergravity theories for a given gravitino
mass $m_{3/2}$ and cosmological constant $V$. We review the known 
bounds which apply to theories with restricted matter content from a new perspective. 
We then extend these results to theories with both hyper and vector multiplets and a 
gauging involving only one generator, for which we show that such a bound exists for 
both $V > 0$ and $V<0$. We finally argue that there is no bound for the same theories 
with a gauging involving two or more generators. These results imply that in N=2 supergravity 
theories metastable de Sitter vacua with $V \ll m^2{\hspace{-4pt}}_{3/2}$ can only arise if at 
least two isometries are gauged, while those with $V \gg m^2{\hspace{-4pt}}_{3/2}$ can 
also arise when a single isometry is gauged.

\end{abstract}

\bigskip

\end{center}
\end{titlepage}

\newpage

\section{Introduction}
\setcounter{equation}{0}

Spontaneous supersymmetry breaking in a vacuum that is at least metastable is 
notoriously difficult to achieve in N=2 supersymmetric theories. This is related to 
the very constrained structure of these theories, with a non-linear sigma-model 
involving target spaces with special geometries and potentials related to the 
gauging of some isometries on these manifolds. In the interesting case of theories 
that can be consistently coupled to gravity, several general results concerning the 
scalar masses in supersymmetry breaking vacua have been obtained, both for rigid 
and local supersymmetry. 
For theories with $n_H$ hypermultiplets and no vector multiplets, it was proven 
in \cite{GRLS} that at least one of the scalars must have square mass $m^2 \le -\, V - \frac 13\, m^2_{3/2}$. 
Similarly, for theories with $n_V$ Abelian vector multiplets and no hypermultiplets, it was shown in \cite{MANY} 
(see also \cite{FTV}) that the lightest scalar must have square mass $m^2 \le - 2 V$. 
These bounds apply to theories with local supersymmetry, with $V$ parametrizing the vacuum 
energy and $m_{3/2}$ denoting the gravitino mass, and imply that de Sitter vacua are necessarily 
unstable in those theories. They also have a non-trivial meaning in the rigid limit, 
in which they reduce to the statement that at least one scalar has $m^2 \le 0$, implying 
that non-supersymmetric vacua cannot be completely stable \cite{JS}. These universal 
results were derived by looking at the averaged sGoldstino mass, for which the dependence 
on the curvature of the scalar manifold turns out to drop out completely, contrary to what happens 
for N=1 theories \cite{GRS1,GRS2,GRS3} (see also \cite{DD,BS,CGRGLPS1,CGRGLPS2}) or 
even N=2 to N=1 truncated theories \cite{CSS1}.\footnote{See \cite{BR,BLR} for similar analyses 
applied to the cases of N=4 and N=8 theories, which are even more constrained and involve fixed 
coset spaces with Planckian curvature as scalar manifolds.}

The aim of this work is to investigate whether there exists a similar bound on the mass
of the lightest scalar in more general N=2 theories involving $n_H$ hypermultiplets and 
$n_V$ vector multiplets. Little is known so far about the systematics of supersymmetry 
breaking in these theories for general $n_H$ and $n_V$. However, the simplest case 
where $n_H=1$ and $n_V=1$ and a single isometry is gauged has been studied in full 
generality in \cite{LSS} for rigid theories and then in \cite{CSS2} for local theories. Exploiting 
the fact that in such a situation it becomes possible to parametrize the scalar geometries of 
both sectors in a much more concrete way in terms of harmonic functions, it was shown that a 
sharp bound on the mass of the lightest scalar emerges in this case too. For theories with local 
supersymmetry, this now approximately reads $m^2 \lsim - \frac 12 V + \frac 14 V^2/m^2_{3/2}$ 
for $V>0$ and is given by similar simple expressions in various ranges of $V<0$, and shows 
that de Sitter vacua can be metastable for sufficiently large cosmological constant. 
For theories with rigid supersymmetry, the corresponding bound is best expressed as $m^2 \le M^2$,
where $M$ denotes the vector mass, and allows non-supersymmetric vacua to 
be metastable. These universal results were derived by reducing the full 
five-dimensional scalar mass matrix to a two-dimensional one by suitably averaging 
over the three physical scalars in the hyper sector and the two scalars in the vector sector. 
It is then clear that this approach does not quite correspond to just looking at the averaged 
sGoldstino mass, but rather exploits to some extent the distinction between the hyper and 
vector sectors. It is then natural to wonder whether a similar bound also persists in theories
with generic $n_H$ and $n_V$. We will prove that this is indeed the case if there is 
only one gauged isometry. However, we will then argue that as soon as there are two 
or more gauged isometries, no universal bound is left, or in other words $m^2 < +\infty$.
In all our analysis we shall focus for simplicity on theories with Abelian gaugings. 
But, it is clear a posteriori that this does not represent a true limitation in the reach 
of our conclusion, since non-Abelian gaugings necessarily involve at least two gauged 
isometries, with which it is already possible to avoid a bound with an Abelian gauging. 

The rest of the paper is organized as follows. In section 2 we briefly review the general
structure of N=2 gauged supergravity theories, focusing on Abelian gaugings. In section 3
and 4 we then review how the known universal bounds on the square mass of the lightest 
scalar arises in theories with only hyper multiplets and only vector multiplets, emphasizing 
the crucial features that allow us to get rid of any dependence on the curvature of the scalar manifold. 
In section 5 we derive a new universal bound for theories with both hyper and vector
multiplets and a single gauged isometry. In section 6 we then study the case of theories
with both hyper and vector multiplets and a more general gauging, and argue that 
as soon as two or more isometries are gauged there is no way to derive any universal
bound that does not depend on the curvature of the scalar manifold, and that by adjusting 
such a curvature at the vacuum point under consideration one can in fact achieve arbitrarily 
large masses for all the scalars. Finally, in section 7 we summarize our main conclusions 
and their implications.

\section{N=2 gauged supergravity}
\setcounter{equation}{0}

Let us consider a general N=2 gauged supergravity theory with $n_H$ hypermultiplets 
and $n_V$ vector multiplets, restricting for simplicity to Abelian symmetries and using Planck units. 
The $4\, n_H$ real scalars $q^u$ from the hypermultiplets span a quaternionic-K\"ahler 
manifold with metric $g_{uv}$ and three almost complex structures $J^x{}^u{\!}_v$ satisfying
$J^x{}^u{\!}_w J^y{}^w{\!}_v = - \delta^{xy} \delta{}^u{\!}_v + \epsilon^{xyz} J^z{}^u{\!}_v$. The 
$n_V$ complex scalars $z^i$ from the vector multiplets span instead a projective special-K\"ahler 
manifold with metric $g_{i \bar \jmath}$ and complex structure $J{}^i{\!}_j = i \delta{}^i{\!}_j$. 
The graviphoton $A_\mu^0$ and the $n_V$ vectors $A_\mu^a$ from the vector multiplets, denoted 
altogether by $A_\mu^A$, have kinetic matrix $\gamma_{AB} = - {\rm Im}\, {\cal N}_{AB}$ and topological 
angles $\theta_{AB} = {\rm Re}\, {\cal N}_{AB}$ in terms of the period matrix ${\cal N}_{AB}$ 
associated with the special-K\"ahler manifold. One can then use these to gauge $n_G \le n_V+1$ 
isometries of the quaternionic-K\"ahler manifold, which are described by triholomorphic 
Killing vectors $k_A^u$ each admitting three Killing prepotentials $P_A^x$ defined by the relations
$\nabla_u P^x_A = - J^{x}_{uv} k_A^v$ and satisfying the equivariance conditions
$J^x_{uv} k^u_A k^v_B =  - \text{\small $\frac 12$} \epsilon^{xyz} P^y_A P^z_B$. 
The scalar and vector kinetic energy is given by \cite{AGF1,AGF2,BW,dWvP,dWLvP,HKLR,S,DFF,ABCDFFM}:
\bea
T = - \text{\small $\frac 14$} \gamma_{\hspace{-1pt}AB} F^A_{\hspace{-1pt}\mu \nu} F^{B\mu \nu} \!
+ \text{\small $\frac 14$} \theta_{\hspace{-1pt}AB} F^A_{\hspace{-1pt}\mu \nu} \tilde F^{B \mu \nu} \!
- \text{\small $\frac 12$} \hspace{1pt} g_{uv}\hspace{1pt} D_\mu q^u D^\mu q^v \!
- g_{i \bar \jmath}\, \partial_\mu z^i \partial^\mu \bar z^{\bar \jmath} \,.
\label{T}
\eea
In this expression $F^A_{\mu \nu} = \partial_\mu A^A_\nu - \partial_\nu A^A_\mu$, 
$\tilde F^A_{\mu \nu} = \frac 12 \epsilon_{\mu \nu \rho \sigma} F^{A \rho \sigma}$ and
$D_\mu q^u = \partial_\mu q^u + k^u_A A^A_\mu$. The scalar potential is instead 
given by
\bea
V = 2 g_{uv} k_A^u k_B^v L^A \bar L^B + g^{i \bar \jmath} f_i^A \bar f_{\bar \jmath}^B P_A^x P_B^x 
- 3 P_A^x P_B^x L^A \bar L^B \,.
\label{V}
\eea
Here $L^A$ denotes a generic covariantly holomorphic symplectic section of the special-K\"ahler manifold
and $f_i^A = \nabla_i L^A$.\footnote{Note that throughout we shall only consider electric gaugings. 
We can do this without loss of generality as we shall not use
special coordinates and therefore we are not restricting ourselves to be in a symplectic frame in 
which a prepotential exists.} Notice that the two $(n_V+1) \times (n_V+1)$ 
matrices $g^{i \bar \jmath} f_i^A \bar f_{\bar \jmath}^B$ and $P_A^x P_B^x$ are both positive definite. 
But since the $A,B$ indices run over $n_V+1$ values, while the indices $i,j$ run only over 
$n_V$ values and the indices $x,y$ only over $3$ values, the first is always singular with 
$1$ null vector and the second is singular whenever $n_V > 3$ with $n_V-3$ null vectors.
Finally, let us also recall that the average gravitino mass is given by
\bea
m^2_{3/2} = P_A^x P_B^x L^A \bar L^B \,.
\eea

The gauge symmetries are generically all spontaneously broken through the VEVs of the scalar 
transformation laws. The latter involve the Killing vectors $k^u_A$, and the order parameters 
of symmetry breaking are described by the matrix of scalar products of these vectors, which 
is recognized to be the vector mass matrix:
\bea
M^2_{AB} = 2 g_{uv} k_A^u k_B^v \,.
\eea
Supersymmetry is also generically completely broken through the VEVs of the hyperini and gaugini
transformation laws. These involve the vectors $N_u^x = |\nabla_u P_A^x L^A|$ and 
$W_i^x = f_i^A P_A^x$ in the two sectors, respectively, and the order parameter of supersymmetry breaking 
is described by the sum of the norms
of these two vectors, which is recognized to be the positive definite part of the potential, namely
$V + 3\hspace{1pt} m^2_{3/2} = \frac 23 g^{uv} N_u^x N_v^x + g^{i \bar \jmath} W_i^x \bar W_{\bar \jmath}^x$.

The quaternionic-K\"ahler manifold describing the hypermultiplet sector has holonomy 
$SU(2) \times SP(2 n_H)$ and a curvature tensor that can be parametrized by a 
four-index tensor $\Sigma_{urvs}$ enjoying some special properties:
\be
R_{urvs} = - \text{\small $\frac 12$} \big(g_{u [v} g_{rs]} + J^x_{ur} J^x_{vs} + J^x_{u[v} J^x_{rs]}\big) 
+ \Sigma_{urvs} \,.
\label{RiemannHyper}
\ee
The tensor $\Sigma_{urvs}$ has the same symmetry properties as the Riemann tensor, but is 
restricted to take the general form $\Sigma_{urvs} = \epsilon_{\Theta \Lambda} \epsilon_{\Pi \Psi}\, {\cal U}_u^{\Theta \theta} 
{\cal U}_r^{\Lambda \lambda} {\cal U}_v^{\Pi \pi} {\cal U}_s^{\Psi \psi} \Sigma_{\theta \hspace{-0.5pt} \lambda \pi \psi}$, 
where ${\cal U}^{\Theta \theta}_u$ denotes the vielbein, $\epsilon_{\Theta \Lambda}$ is the antisymmetric symbol of $SU(2)$ 
and $\Sigma_{\theta \hspace{-0.5pt} \lambda \pi \psi}$ is an arbitrary completely symmetric $SP(2n_H)$ tensor. 
As a result of its very special form, $\Sigma_{urvs}$ gives a very restricted, specific 
contribution to the curvature. Firstly, it does not contribute to the contractions defining the Ricci and 
the scalar curvature, which are thus completely fixed and given by
\be
R_{uv} = - (n_H+2) g_{uv} \,,\;\; R = - 4 n_H (n_H + 2) \, .
\label{RicciScalarHyper}
\ee
Secondly, it also does not contribute to the completely symmetric part of the Riemannian curvature
contracted with the sum of the product of two complex structures. Indeed, the complex structure can 
be rewritten as $J^x_{uv} = i \sigma^x_{\Theta \Lambda} c_{\theta \lambda}\, {\cal U}_u^{\Theta \theta} {\cal U}_v^{\Lambda \lambda}$, 
where $c_{\theta \lambda}$ denotes the antisymmetric symbol of $SP(2n_H)$, and using the property 
$\sigma^x_{\Theta \Lambda} \sigma^x_{\Pi \Psi} = - 2\hspace{1pt} \epsilon_{\Theta(\Pi} \epsilon_{\Lambda \Psi)}$ 
one finds that the following quantity is completely fixed:
\bea
R_{(urvs} J^x{}^r{\!}_{p} J^x{}^s{\!}_{q)} = - 3\hspace{1pt} g_{(uv} g_{pq)} \,.  
\label{RiemannPropHyper}
\eea

The special-K\"ahler manifold describing the vector multiplet sector has instead a curvature tensor 
that can be entirely characterized by a three-index tensor $C_{ijk}$ enjoying some special properties:
\bea
R_{i \bar \jmath p \bar q} = g_{i \bar \jmath} g_{p \bar q} + g_{i \bar q} g_{p \bar \jmath} 
- C_{i p r} g^{r \bar s} \bar C_{\bar \jmath \bar q \bar s} \,.
\label{RiemannVector}
\eea
The tensor $C_{ijk}$ must be completely symmetric and covariantly holomorphic, but is otherwise arbitrary.
It also controls the second covariant derivatives of the symplectic section, which read:
\be
\nabla_i f_j^A = C_{ijk} \bar f^{k A} \,,\;\; \nabla_{i} \bar f_{\bar \jmath}^A = g_{i \bar \jmath} \bar L^A \,.
\ee

A supersymmetry breaking vacuum is generically associated to a point on the scalar manifold at 
which $V > - 3\hspace{3pt}m^2_{3/2}$ and $V' = 0$, and the mass matrix for scalar fluctuations 
is then related to the value of the Hessian matrix $V''$ at such a point. To explore the existence 
of possible obstructions to making all the scalars arbitrarily heavy by adjusting the parameters 
of the theory, one may choose an arbitrary point on the scalar manifold with fixed values of $V$ and 
$m^2_{3/2}$ and impose the stationarity conditions. The latter are then viewed as restrictions on the parameters 
of the theory, ensuring that the point under consideration is indeed a good vacuum. One then computes 
the scalar mass matrix and checks whether its eigenvalues can be made arbitrarily large or not whilst obeying 
the previous constraints. The general strategy to look for a non-trivial bound is then to study the 
scalar mass matrix along the particular directions in field space defined by the shift vectors 
$N_u^x = |\nabla_u P_A^x L^A|$ and $W_i^x = f_i^A P_A^x$, which determine the sGoldstino directions 
and are well defined under the assumption that supersymmetry is spontaneously broken both in the hyper 
and the vector sectors. By suitably averaging over all such directions, one may finally derive a single 
universal bound in units of $m^2_{3/2}$, which will depend on the following parameter controlling the 
cosmological constant $V$:
\be
\epsilon = \frac {V}{m^2_{3/2}} \,.
\label{eq:ep}
\ee
Recall, finally, that vacuum metastability requires scalar masses to satisfy $m^2 > 0$ in de Sitter vacua 
with $\epsilon \in (0,+\infty)$ and $m^2 > \frac 34 \epsilon\hspace{1pt} m^2_{3/2}$ in anti-de Sitter vacua with 
$\epsilon \in (-3,0)$.

\section{Only hypers}
\setcounter{equation}{0}

Let us first briefly review the case of theories with $n_H$ hypers and no vectors with a gauging involving 
just the graviphoton, following \cite{GRLS}. In this case, $L^0$ is a constant. We can then define 
$k^u = k^u_0 |L^0|$ and $P^x = P^x_0 |L^0|$. In this way, the potential reads:
\bea
\b\b V = 2\hspace{1pt} k^w k_w - 3\hspace{1pt} P^z P^z \,.
\eea
The stationarity condition is obtained by computing the first covariant derivative and setting it to zero.
This yields:
\bea
\b\b 4\hspace{1pt} k^w \nabla_u k_w - 6\hspace{1pt} P^z \nabla_u P^z = 0 \,.
\label{StatHyper}
\eea
The unnormalized scalar mass matrix is then defined by the second covariant derivative evaluated at the stationary 
point under consideration and reads:
\bea
m^2_{uv} = - 4 \big(R_{u r v s} k^r k^s - \nabla_u k^w \nabla_v k_w\big)
- 6 \big(P^z \nabla_u \nabla_v P^z + \nabla_u P^z \nabla_v P^z\big) \,.
\eea
The gravitino mass is instead given by:
\be
m^2_{3/2} = P^z P^z \,.
\ee

One may now look at the mass matrix along the special set of vectors $N_u^x = \nabla_u P^x$
defining the sGoldstino directions, or equivalently
\bea
n^{ux} = \frac {\nabla^u P^x}{\sqrt{k^w k_w}} = - J^x{}^u{}_v \frac {k^v}{\sqrt{k^w k_w}}\,.
\eea
These are orthonormal with respect to the metric and satisfy $g_{uv} n^{ux} n^{vy} = \delta^{xy}$.
One may then consider the following quantity, corresponding to the physical average 
sGoldstino mass:
\bea
m^2_{\rm bound} \equiv \text{\small $\frac 13$} m^2_{uv} n^{xu} n^{xv} \,.
\eea
This quantity $m^2_{\rm bound}$ represents by construction an upper bound on the square mass 
of the lightest scalar, and also a lower bound on that of the heaviest. Indeed, for each fixed $x=1,2,3$ 
the quantity $m^2_{uv} n^{xu} n^{xv}$ (no sum over $x$) is a normalized combination of the eigenvalues 
of the matrix $m^2_{uv}$ yielding its value along the unit vector $n^x_u$, which manifestly provides such type 
of bounds. The quantity $\text{\small $\frac 13$} m^2_{uv} n^{xu} n^{xv}$ (sum over $x$) then corresponds 
to the average of the above quantities over $x=1,2,3$, and thus also provides such type of bounds.

To evaluate more concretely the form of $m^2_{\rm bound}$, let us parametrize the potential $V$ in 
terms of the gravitino mass $m^2_{3/2}$ as 
\be
V = (x - 3) m^2_{3/2} \,,
\ee
with
\be
x = \frac {2 k^w k_w}{P^z P^z} \,.
\ee
By using the stationarity condition (\ref{StatHyper}) (which is easily shown to imply
that $\nabla_u k^w n^{ux} = \frac 12 (\delta^{xy} J^z{}^w{\!}_u + \epsilon^{xyz} \delta^w{\!\!\!}_u) P^y n^{uz}$)
and the special property (\ref{RiemannPropHyper}) for the curvature, we then see that all the 
dependence on the curvature drops out and $m^2_{\rm bound}$ is found to be given by the following 
universal value:
\bea
m^2_{\rm bound} = \text{\small $\frac 13$} \big(8 - 3\hspace{1pt}x \big) m^2_{3/2}\,.
\eea

We can finally rewrite the above result in terms of the dimensionless parameter $\epsilon$
defined in (\ref{eq:ep}), which controls the cosmological constant. One simply has $x = 3+\epsilon$, 
and therefore:
\bea
m^2_{\rm bound} = - \text{\small $\frac 13$} \big(1 + 3\hspace{1pt} \epsilon \big) m^2_{3/2}\,.
\eea
In terms of $V$ and $m^2_{3/2}$, this finally means:
\bea
m^2_{\rm bound} = -\, V - \text{\small $\frac 13$} m^2_{3/2} \,.
\eea
This result shows that within this class of N=2 theories, de Sitter vacua are unavoidably unstable for any 
positive value $V > 0$ of the cosmological constant, while anti-de Sitter vacua can be metastable 
only for sufficiently negative values $V \in (-3\hspace{1pt} m^2_{3/2},-\frac 4{21}\hspace{1pt} m^2_{3/2})$ 
of the cosmological constant \cite{GRLS}.

\section{Only vectors}
\setcounter{equation}{0}

Let us next briefly review also the case of theories with $n_V$ vector multiplets with constant Fayet-Iliopoulos 
terms and no hypers, following \cite{FTV}. In this case, the equivariance conditions force all the $P^x_A$, seen as 
$n_V$ tridimensional vectors, to be parallel. We can then write $P^x_A = \xi_A v^x$ with $v^x v^x = 1$, and 
define $L = \xi_A L^A$ and $f_i = \xi_A f_i^A$. In this way, the potential becomes
\bea
\b\b V = \bar f^k f_k - 3\hspace{1pt} |L|^2 \,,
\eea
and the stationarity condition reads
\bea
\b\b C_{ikl} \bar f^{k} \bar f^{l} - 2\hspace{1pt} f_i \bar L = 0 \,.
\label{StatVector}
\eea
The Hermitian block of the unnormalized scalar mass matrix is then defined by the second mixed 
covariant derivatives evaluated at the stationary point under consideration and reads:\footnote{The 
off-diagonal complex block $m^2_{ij}$ will play no role in the following.}
\bea
m^2_{i \bar \jmath} = - 2\hspace{1pt} R_{i \bar \jmath p \bar q} \bar f^{p} f^{\bar q}
+ 2\hspace{1pt} g_{i \bar \jmath} \big(\bar f^k f_k - |L|^2 \big) \,.
\eea
The gravitino mass is finally given by
\be
m^2_{3/2} = |L|^2 \,.
\ee

One may now look at the mass matrix along the special vector $W_i = f_i$ defining the sGoldstino 
direction, or equivalently
\bea
w^i = \frac {\bar f^i}{\sqrt{\bar f^k f_k}} \,.
\eea
This is normalized with respect to the metric and satisfies $g_{i \bar \jmath} w^i \bar w^{\bar \jmath} = 1$.
One may then consider the following quantity, corresponding to the physical average 
sGoldstino mass:
\bea
m^2_{\rm bound} \equiv m^2_{i \bar \jmath} w^i \bar w^{\bar \jmath} \,.
\eea
This quantity $m^2_{\rm bound}$ represents by construction an upper bound on the square mass of the 
lightest scalar, and also a lower bound on that of the heaviest. To see this, let us switch to a real notation 
with $I = i,\bar \imath$ and introduce the two unit vectors $w_+^I = \text{\small $\frac 1{\sqrt{2}}$} (w^i,\bar w^{\bar \jmath})$, 
$w_-^I = \text{\small $\frac i{\sqrt{2}}$}(w^i,-\bar w^{\bar \jmath})$, so that
$m^2_{\rm bound} = \text{\small $\frac 12$} m^2_{I \bar J} w_s^I w_s^{\bar J}$ with $s = \pm$.
One may then argue exactly as in the previous section. 
For each fixed $s = \pm$, the quantity $m^2_{I \bar J} w_s^I w_s^{\bar J}$ (no sum over $s$) is a normalized 
combination of the eigenvalues of the matrix $m^2_{I \bar J}$ yielding its value along the unit vector $w^I_s$, 
which manifestly provides such type of bounds. The quantity $\text{\small $\frac 12$} m^2_{I \bar J} w_s^I w_s^{\bar J}$ 
(sum over $s$) then gives the average of these quantities over $s=\pm$, and thus also yields
such type of bounds.

To explicitly evaluate the form of $m^2_{\rm bound}$, let us parametrize the potential $V$ in 
terms of the gravitino mass $m^2_{3/2}$ as 
\be\label{eq:Vh}
V = (y - 3) m^2_{3/2} \,,
\ee
with
\be
y = \frac {\bar f^k f_k}{|L|^2} \,.
\ee
By using the stationarity condition (\ref{StatVector}) and the form (\ref{RiemannVector}) for the curvature, 
we then see that once again all the dependence on the curvature drops out and $m^2_{\rm bound}$ is 
found to be given by the following universal value:
\bea
m^2_{\rm bound} = 2 \big(3 - y \big) m^2_{3/2}\,.
\eea

We can again rewrite the above result in terms of the dimensionless parameter $\epsilon$
defined in (\ref{eq:ep}), which controls the cosmological constant. One simply has $y = 3 + \epsilon$, 
and therefore:
\bea
m^2_{\rm bound} = - 2 \hspace{1pt}\epsilon\hspace{1pt}  m^2_{3/2}\,.
\eea
In terms of $V$ and $m^2_{3/2}$, this finally means:
\bea
m^2_{\rm bound} = - 2\, V \,.
\eea
This result shows that within this class of N=2 theories de Sitter vacua are unavoidably unstable for any 
positive value $V > 0$ of the cosmological constant, while anti-de Sitter vacua can be metastable 
for any negative value $V \in (-3\hspace{1pt} m^2_{3/2}, 0)$ of the cosmological constant \cite{MANY}.

\section{Hypers and vectors with one gauging}
\setcounter{equation}{0}

Let us now study what happens in the more general case of theories with $n_H$ hypers 
and $n_V$ vectors with a gauging involving both the vectors and the 
graviphoton but only $1$ isometry. It turns out that this case is still simple enough 
to allow the derivation of a universal bound generalizing that derived in \cite{CSS2}. 
In this situation we have $k^u_A = \xi_A k^u$ and $P^x_A = \xi_A P^x$,
and we can define $L = \xi_A L^A$ and $f_i = \xi_A f^A_i$. The potential then reads
\bea
V = 2\hspace{1pt} k^w k_w |L|^2 + \big(\bar f^k f_k  - 3\hspace{1pt} |L|^2 \big) P^z P^z\,,
\eea
and the stationarity conditions are given by
\bea
\b\b 4\hspace{1pt} k^w \nabla_u k_w |L|^2 + 2\big(\bar f^k f_k - 3\hspace{1pt} |L|^2\big) P^z \nabla_u P^z = 0 \,, 
\label{Stat1HyperVector}\\[1mm]
\b\b C_{ikl} \bar f^{k} \bar f^{l} P^z P^z + 2\big(k^w k_w - P^z P^z \big) f_i \bar L = 0 \,.
\label{Stat2HyperVector}
\eea
The relevant blocks of the unnormalized scalar mass matrix are then found to be:
\bea
\b\b m^2_{uv\!} = -\, 4 \big(R_{u r v s} k^r k^s - \nabla_u k^w \nabla_v k_w\big) |L|^2 \nn \\[1mm]
\b\b \hspace{35pt} +\, 2 \big(\bar f^k f_k - 3 |L|^2\big) \big(P^z \nabla_u \nabla_v P^z + \nabla_u P^z \nabla_v P^z \big) \,, \\[1mm]
\b\b m^2_{\hspace{1pt} i \bar \jmath \hspace{1pt}} = - 2\hspace{1pt} R_{i \bar \jmath p \bar q} \bar f^{p} f^{\bar q} P^z P^z
+ 2\hspace{1pt} f_i \bar f_{\bar \jmath}\, k^w k_w \nn \\[0.5mm]
\b\b \hspace{35pt} +\, 2\hspace{1pt} g_{i \bar \jmath} |L|^2\, k^w k_w 
+ 2\hspace{1pt} g_{i \bar \jmath} \big(\bar f^k f_k - |L|^2 \big) P^z P^z \,, \\[1mm]
\b\b m^2_{u i} = 4\hspace{1pt} k^w \nabla_u k_w f_i \bar L + 2 \big(C_{ikl} \bar f^{k} \bar f^{l}  - 2\hspace{1pt} f_i \bar L\big) P^z \nabla_u P^z  \,.
\eea
The gravitino mass is instead given by
\be
m^2_{3/2} = P^z P^z |L|^2 \,.
\ee

One may at this point look at the mass matrix along the special sets of vectors $N_u^x = \nabla_u P^x |L|$ and $W_i = f_i$ 
defining the sGoldstino directions in the hyper and vector subsectors, corresponding to
\bea
n^{ux} = \frac {\nabla^u P^x}{\sqrt{k^w k_w}} = - J^x{}^u{}_v \frac {k^v}{\sqrt{k^w k_w}}\,,\;\; 
w^i = \frac {\bar f^i}{\sqrt{\bar f^k f_k}} \,.
\eea
These are orthonormal with respect to the metric and satisfy $g_{uv} n^{ux} n^{vy} = \delta^{xy}$ and 
$g_{i \bar \jmath} w^i \bar w^{\bar \jmath} = 1$. Generalizing the approach of \cite{LSS,CSS2}, one 
may then consider the following $2 \times 2$ matrix, 
obtained by averaging over the sGoldstino directions separately in the two sectors:
\bea
m^2_{\rm avr} \equiv \left(\,\begin{matrix}
m^2_{\rm hh} \!&\! m^2_{\rm hv} \smallskip\ \\
m^2_{\rm hv}  \!&\! m^2_{\rm vv} \\
\end{matrix} \! \right) \,,
\label{m2avr}
\eea
where 
\be
m^2_{\rm hh} \equiv \text{\small $\frac 13$} m^2_{u v} n^{ux} n^{vx} \,,\;\;
m^2_{\rm vv} \equiv m^2_{i \bar \jmath} w^i \bar w^{\bar \jmath} \,,\;\;
m^2_{\rm hv} \equiv  \sqrt{\text{\small $\frac 13$} m^2_{u i} n^{ux} w^i m^2_{v \bar \jmath} n^{vx} \bar w^{\bar \jmath}}\,.
\ee
The two eigenvalues of this averaged matrix are:
\be
m^2_{\rm \pm} = \text{\small $\frac 12$} \big(m^2_{\rm hh} + m^2_{\rm vv}\big)
\pm \sqrt{\text{\small $\frac 14$} \big(m^2_{\rm hh} - m^2_{\rm vv} \big)^2 + m^4_{\rm hv}} \,.
\label{m2pm}
\ee
These quantities $m^2_-$ and $m^2_+$ yield by construction an upper bound on the square mass of the 
lightest scalar and a lower bound on that of the heaviest, respectively. This can be proven through some 
simple linear algebra, by switching to a real notation and proceeding as follows. One starts by constructing 
the $5 \times 5$ restriction of the mass matrix onto the vector space spanned by the $3$ unit vectors $n^{ux}$ 
in the hyper sector and the $2$ independent unit vectors $w_\pm^I$ associated to the complex $w^i$ in the 
vector sector. This involves a $3 \times 3$ diagonal hyper-hyper block, a $2 \times 2$ diagonal vector-vector
block, and a $3 \times 2$ off-diagonal hyper-vector block. One then considers the two $5 \times 5$ matrices 
obtained by subtracting from this restricted mass matrix the unit matrix multiplied respectively by the smallest 
and the largest of its eigenvalues. By construction these two matrices must be respectively positive and 
negative definite. One finally shows that this implies that the minimal and maximal eigenvalues of the 
restricted mass matrix, and thus also those of the full mass matrix, must be smaller than the minimal 
eigenvalue of the $2 \times 2$ averaged mass matrix (\ref{m2avr}) and larger than the maximal one,
respectively.

To evaluate more concretely the form of $m^2_{\rm hh}$, $m^2_{\rm vv}$, $m^2_{\rm hv}$, let 
us parametrize the potential $V$ in terms of the gravitino mass $m^2_{3/2}$ as 
\be
V = (x + y - 3) m^2_{3/2} \,,
\ee
with
\be
x = \frac {2 k^w k_w}{P^z P^z} \,,\;\; y = \frac {\bar f^k f_k}{|L|^2} \,.
\ee
We can now simplify the averaged masses by using the stationarity conditions 
(\ref{Stat1HyperVector}) (which implies $\nabla_u k^w n^{ux} = \frac 12 [\frac 13(3 - y) \delta^{xy} J^z{}^w{\!}_u 
+ (1- y) \epsilon^{xyz} \delta^w{\!\!\!}_u] P^y n^{uz}$) and (\ref{Stat2HyperVector}), and the relations (\ref{RiemannPropHyper}) and
(\ref{RiemannVector}) for the curvatures. Proceeding as before, we see that all the dependence on the curvature 
drops out, as in the previous two cases, and $m^2_{\rm hh}$, $m^2_{\rm vv}$, $m^2_{\rm hv}$ 
are found to be given by the following universal values:
\bea
\b\b m^2_{\rm hh} = \text{\small $\frac 13$} \big(y - 1\big) \big(3x + 4y - 8\big) m_{3/2}^2 \,,\\[1.5mm]
\b\b m^2_{\rm vv} = \big(x - 2) \big(2x + y - 3\big) m_{3/2}^2 \,, \\[-0.5mm]
\b\b m^2_{\rm hv} = \sqrt{\text{\small $\frac 23$}} \sqrt{xy\raisebox{6pt}{$$}}\,\big(x+y-3\big) m^2_{3/2} \,.
\eea

We can now rewrite the above results in terms of the parameter $\epsilon$ defined 
in (\ref{eq:ep}), which controls the cosmological constant, and an angle $\theta$ 
parametrizing the relative importance of the contributions of the two sectors to 
supersymmetry breaking:
\be
\tan^2 \theta = \frac yx \,.
\ee
One then has $x = (3 + \epsilon) \cos^2 \theta$ and $y = (3 + \epsilon) \sin^2 \theta$, and the entries
of the averaged mass matrix can be rewritten in the following form:
\bea
\b\b m^2_{\rm hh} = \Big[\text{\small $\frac 13$} \big((3 + \epsilon) \cos^2\! \theta - (2 + \epsilon) \big) 
\big((3 + \epsilon) \cos^2\! \theta - 4 (1 + \epsilon) \big) \Big]m^2_{3/2}
\label{m2hh} \,, \\
\b\b m^2_{\rm vv} = \Big[\big((3 + \epsilon) \cos^2\! \theta - 2\big) \big((3 + \epsilon) \cos^2\! \theta + \epsilon\big) \Big] m^2_{3/2} 
\label{m2vv}\,, \\[-1mm]
\b\b m^2_{\rm hv} = \Big[\text{\small $\sqrt{\frac 23}$}\hspace{1pt} \epsilon \hspace{1pt} (3 + \epsilon) \cos \theta \sin \theta \Big] m^2_{3/2} 
\label{m2hv}\,.
\eea
These are now recognized to be exactly the same results that were obtained in \cite{CSS2} for the special case 
of theories with $n_H = 1$ and $n_V = 1$ based on a single gauge symmetry. As a consequence, all the results 
derived in \cite{CSS2} generalize to any theory with hypers and vectors but a single gauge symmetry. In particular, 
the main features of the two eigenvalues $m^2_\pm$ were shown to be the following. In the limit  $\theta \to 0$, 
in which the hyper sector dominates supersymmetry breaking, one finds 
$m^2_\pm \to {}^{\rm max}{\hspace{-15pt}}_{\rm min}\,\big\{\!\!-\! \frac 13 (1 + 3 \hspace{1pt} \epsilon) \, m_{3/2}^2, 
(1 + \epsilon) (3 + 2\hspace{1pt} \epsilon)\, m_{3/2}^2 \big\}$. One of the eigenvalues thus corresponds to the value 
of $m^2_{\rm bound}$ found for theories with just hypers, while the other corresponds to the mass of a combination of scalars 
from the vector sector. Depending on the situation, either of the two can be the smallest or the largest one. 
In the limit  $\theta \to \frac {\pi}2$, in which the vector sector dominates supersymmetry breaking, one finds 
$m^2_\pm \to {}^{\rm max}{\hspace{-15pt}}_{\rm min}\,\big\{\!\!-\! 2\hspace{1pt} \epsilon \, m_{3/2}^2, 
\frac 43 (1 + \epsilon) (2 + \epsilon) \, m_{3/2}^2 \big\}$. Again, depending on the situation, either of the two 
can be the smallest or the largest one. Finally, when $\theta$ has an intermediate value and both sectors 
contribute comparably to supersymmetry breaking, one finds a much more complicated result. For any possible 
value for $\epsilon$, one may then scan over the possible values of $\theta$ and determine the maximal and 
minimal values of $m^2_-$ and $m^2_+$ taken on they own, namely:
\bea
\b\b m^2_{\rm up} \equiv \max_{\theta} \big\{m^2_-\big\} \,, \\
\b\b m^2_{\rm low} \equiv \min_{\theta} \big\{m^2_+\big\} \,.
\eea
The quantities $m^2_{\rm up}$ and $m^2_{\rm low}$ still represent by construction an upper bound to the square mass 
of the lightest scalar and a lower bound to that of the heaviest. Their precise values as functions of $\epsilon$ can only be 
computed numerically. However, their behavior is mainly determined by the fact that when changing the value of $\epsilon$
in the range $(-3,+\infty)$, the optimal value for $\theta$ that extremizes $m^2_\pm$ switches among the three 
situations in which one, the other or both sectors dominate supersymmetry breaking. Using this observation, 
one can then derive the following approximate analytic expressions for $m^2_{\rm up}$ and $m^2_{\rm low}$, 
which are constructed in a such a way that they reproduce the correct asymptotic behaviors for small and large $\epsilon$ 
and define a bound that is still valid but no-longer saturable:
\bea
\b\b m^2_{\rm up} \simeq \left\{\!\!
\begin{array}{ll} 
- \text{\small $\frac 13$} \big(1+ 3\hspace{1pt} \epsilon\big) m^2_{3/2} \,,& \epsilon \in (-3, - \frac {9 + \sqrt{21}}6]\\[2mm]
\big(1+\epsilon\big)\big(3 + 2\hspace{1pt} \epsilon \big) m^2_{3/2} \,,& \epsilon \in [- \frac {9 + \sqrt{21}}6,-1]\\[2mm]
\text{\small $\frac 43$} \big(1+ \epsilon\big)\big(2 + \epsilon \big) m^2_{3/2} \,,& \epsilon \in [-1,-\frac 12]\\[2mm]
- 2\hspace{1pt} \epsilon\hspace{1pt} \hspace{1pt} m^2_{3/2} \,,& \epsilon \in [-\frac 12,0]\\[2mm]
- \text{\small $\frac 14$} \epsilon \big(2 - \epsilon\big) m^2_{3/2} \,,& \epsilon \in [0,+\infty)
\end{array} 
\right. \,,
\label{Mupapprox}
\eea
\vskip -15pt
\bea
\b\b m^2_{\rm low\!\!} \simeq \left\{\!\!
\begin{array}{ll} 
\big(1+\epsilon\big)\big(3 + 2\hspace{1pt} \epsilon \big) m^2_{3/2} \,,& \epsilon \in (-3, - \frac {9 + \sqrt{21}}6]\\[2mm]
- \text{\small $\frac 13$} \big(1+ 3\hspace{1pt} \epsilon\big) m^2_{3/2} \,,& \epsilon \in [- \frac {9 + \sqrt{21}}6,- \frac {9 - \sqrt{21}}6]\\[2mm]
\big(1+\epsilon\big)\big(3 + 2\hspace{1pt} \epsilon \big) m^2_{3/2} \,,& \epsilon \in [- \frac {9 - \sqrt{21}}6,-0.71]\\[2mm]
- \text{\small $\frac 12$} \epsilon \big(1 - 0.44\hspace{1pt} \epsilon\big) m^2_{3/2} \,,& \epsilon \in [-0.71,0] \\[2mm]
\text{\small $\frac 32$} \epsilon \big(1 + 0.70\hspace{1pt} \epsilon\big) m^2_{3/2} \,,& \epsilon \in [0,+\infty) 
\end{array}
\right. \,.
\label{Mlowapprox}
\eea
In terms of $V$ and $m^2_{3/2}$, this finally means that for $V < 0$ one has various branches with simple but 
different functional behaviors that always stay above the stability bound $\frac 34 V$, while for $V>0$ one 
has the following approximate behavior:
\bea
\b\b m^2_{\rm up} \simeq 
-\,\text{\small $\frac 12$} V + \text{\small $\frac 14$} \frac {V^2}{m^2_{3/2}} \,, \\
\b\b m^2_{\rm low} \simeq  \text{\small $\frac 32$} V + 1.05 \frac {V^2}{m^2_{3/2}}\,.
\eea
These results, which are depicted in figure 1, show that within this class of N=2 theories de Sitter vacua 
can be metastable, but only for sufficiently large positive values $V \gsim 2\hspace{1pt} m^2_{3/2}$ 
(or more precisely $V > 2.17\hspace{1pt} m^2_{3/2}$ according to a numerical analysis) of 
the cosmological constant, while anti-de Sitter vacua can be metastable for any negative value 
$V \in (-3\hspace{1pt} m^2_{3/2},0)$ of the cosmological constant.

\begin{figure}[h]
\vskip 20pt
\begin{center}
\includegraphics[width=0.6\textwidth]{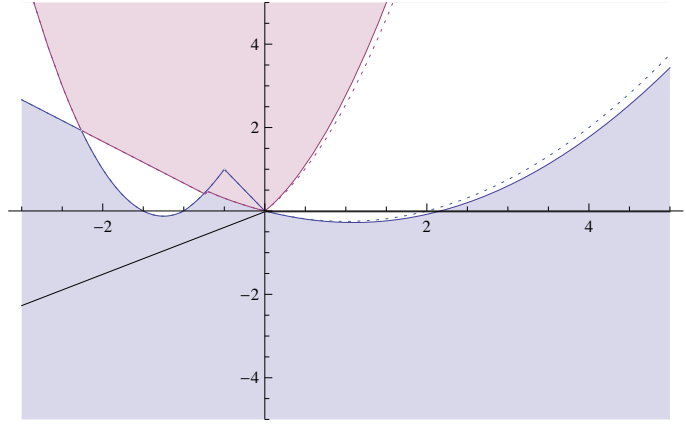}
\caption{Plot of the exact expressions determined numerically (solid curves) and the approximate analytical expressions 
in (\ref{Mupapprox}) and (\ref{Mlowapprox}) (dotted curves) for the upper and lower bounds $m^2_{\rm up}$ (blue) and 
$m^2_{\rm low}$ (red) as functions of $\epsilon$. The two shaded areas (blue and red regions) delimit the ranges in 
which the smallest and the largest mass eigenvalues are allowed to lie. The metastability bound is also shown (solid 
black lines).}
\end{center}
\end{figure}

\section{Hypers and vectors with several gaugings}
\setcounter{equation}{0}

Let us finally try to see what happens in the general case of theories with $n_H$ hypers 
and $n_V$ vectors with a gauging involving both the vectors and the 
graviphoton and a generic number $n_G$ of isometries. 
In this general situation, there is no way to avoid the explicit appearance of some indices 
labeling the gauged isometries. At most, one may switch from the indices $A,B$ 
running over the $n_V+1$ vector fields to new indices $\alpha,\beta$ running only over the 
$n_G$ gauged isometries by first rewriting $k^u_A = \xi_A^\alpha k_\alpha^u$ 
and $P^x_A = \xi_A^\alpha P_\alpha^x$, and then defining $L^\alpha = \xi_A^\alpha L^A$ and 
$f_i^\alpha = \xi_A^\alpha f^A_i$. The potential then reads
\bea
V = 2\hspace{1pt} k_\alpha^w k_{\beta w} L^\alpha \bar L^\beta + \bar f^{k\alpha} f_k^\beta P_\alpha^z P_\beta^z 
- 3\hspace{1pt} P_\alpha^z P_\beta^z L^\alpha \bar L^\beta \,,
\eea
and the stationarity conditions are given by
\bea
\b\b 4\hspace{1pt} k_\alpha^w \nabla_u k_{\beta w} L^\alpha \bar L^\beta 
+ 2\hspace{1pt} \big(\bar f^{k\alpha} f_k^\beta - 3 L^\alpha \bar L^\beta\big) P_{(\alpha}^z \nabla_u P_{\beta)}^z   = 0 \,, 
\label{Stat1HyperVectorGen}\\[1mm]
\b\b C_{ikl} \bar f^{\alpha k} \bar f^{\beta l} P_\alpha^z P_\beta^z + 2\big(k_\alpha^w k_{\beta w} - P_\alpha^z P_\beta^z \big) 
f_i^\alpha \bar L^\beta = 0 \,.
\label{Stat2HyperVectorGen}
\eea
The relevant blocks of the unnormalized scalar mass matrix are then found to be
\bea
\b\b m^2_{uv\!} = -\, 4 \big(R_{u r v s} k_\alpha^r k_\beta^s  - \nabla_u k_\alpha^w \nabla_v k_{\beta w}\big) L^\alpha \bar L^\beta \nn \\[1mm]
\b\b \hspace{35pt} +\, 2 \big(\bar f^{\alpha k} f_k^\beta - 3 L^\alpha \bar L^\beta \big) \big(P_{(\alpha}^z \nabla_u \nabla_v P_{\beta)}^z 
+ \nabla_u P_{(\alpha}^z \nabla_v P_{\beta)}^z \big) \,, \\[1mm]
\b\b m^2_{\hspace{1pt} i \bar \jmath \hspace{1pt}} = - 2\hspace{1pt} R_{i \bar \jmath p \bar q} \bar f^{\alpha p} f^{\beta\bar q} P_\alpha^x P_\beta^x
+ 2\hspace{1pt} f^\alpha_i \bar f^\beta_{\bar \jmath} k_\alpha^w k_{\beta w} \nn \\[0.5mm]
\b\b \hspace{35pt} +\, 2\hspace{1pt} g_{i \bar \jmath} L^\alpha \bar L^\beta k_\alpha^w k_{\beta w} 
+ 2\hspace{1pt} g_{i \bar \jmath} \big(\bar f^{\alpha k} f_k^\beta - L^\alpha \bar L^\beta \big) P_\alpha^z P_\beta^z \,, \\[1mm]
\b\b m^2_{u i} = 4\hspace{1pt} k_\alpha^w \nabla_u k_{\beta w} f_i^\alpha \bar L^\beta 
+ 2 \big(C_{ikl} \bar f^{\alpha k} \bar f^{\beta l}  - 2\hspace{1pt} f_i^\alpha \bar L^\beta\big) P_{(\alpha}^z \nabla_u P_{\beta)}^z \,.
\eea
The gravitino mass is finally given by
\be
m^2_{3/2} = P_\alpha^z P_\beta^z L^\alpha \bar L^\beta \,.
\ee

One may now try to look at the mass matrix along some special sets of vectors 
defining the sGoldstino directions in the hyper and vector sectors. The natural 
candidates for these are given by the vectors 
$N_u^x = |\nabla_u P_\alpha^x L^\alpha|$ and $W_i^x = f_i^\alpha P_\alpha^x$ 
controlling the shifts of the hyperini and the gaugini under supersymmetry transformations. 
The $3$ vectors $N^{ux}$ are orthogonal and satisfy $g_{uv} N^{ux} N^{vy} = c^{-2} \hspace{1pt} \delta^{xy}$ 
with $c = |k_\alpha^w k_{\beta w} L^\alpha \bar L^\beta|^{-1/2}$. One may then simply rescale the $N^{ux}$ 
to define $3$ orthonormal vectors $n^{ux} = c\hspace{1pt} N^{ux}$.
The $3$ vectors $W^{ix}$ are not orthogonal and instead  satisfy $g_{i \bar \jmath} \bar W^{ix} W^{\bar \jmath y} = d^{xy}$ 
with $d^{xy} = P^x_\alpha P^y_{\beta} (\bar f^{\alpha k} f^\beta_k)$. Moreover, one finds that the $3 \times 3$
matrix $d^{xy}$ has rank $r = 3$ only if $n_G \ge 3$ and $n_V \ge 3$. This matrix has rank $2$ when 
$n_G = 2$ and $n_V \ge 2$, or $n_G = 3$ and $n_V = 2$, and rank $1$ when $n_G =1$ and $n_V \ge 1$ 
(which includes the case studied in the previous section), or $n_G = 2$ and $n_V =1$. 
One may then take suitable linear combinations of the $\bar W^{ix}$ to define $r$ independent orthonormal vectors 
$w^{ix'} = c^{x'}_x \bar W^{ix}$, with $x'=1,..,r$ and $c^{x'}_x$ such that $c^{x'}_x \bar c^{y'}_y d^{xy} = \delta^{x' \hspace{-1pt}y'}$. 
Summarizing, we may thus consider the following two sets of $3$ and $r$ vectors in the two sectors:
\bea
n^{ux} = c\, |\nabla^u P^x_\alpha L^\alpha| = - c\, J^x{}^u{}_v |k_\alpha^v L^\alpha| \,,\;\; 
w^{ix'} = c^{x'}_x \bar f^{\alpha i} P_\alpha^x \,.
\eea
These now satisfy $g_{uv} n^{ux} n^{vy} = \delta^{xy}$ and $g_{i \bar \jmath} w^{ix'} \bar w^{\bar \jmath y'} = \delta^{x' \hspace{-1pt} y'}$.
We may then try to proceed as in the previous section and define a $2 \times 2$ matrix by averaging over these two 
special sets of directions within each of the two sectors, with entries given by:
\bea \label{tmm}
\b\b m^2_{\rm hh} \equiv \text{\small $\frac 13$} m^2_{u v} n^{ux} n^{vx} \,,\;\;
m^2_{\rm vv} \equiv \text{\small $\frac 1r$} m^2_{i \bar \jmath} w^{ix'\!} \bar w^{\bar \jmath x'\!} \,,\\
\b\b m^2_{\rm hv} \equiv  \sqrt{\text{\small $\frac 1{3\hspace{1pt} r}$} m^2_{u i} n^{ux} w^{ix'\!} m^2_{v \bar \jmath} n^{vx} \bar w^{\bar \jmath x'\!}}\,.
\eea
However, it turns out that this no longer allows us to eliminate all the dependence on the curvature, and 
therefore no universal bound emerges in this general case. To see how this comes about, let us focus on 
the terms in the mass matrix that may a priori depend on $\Sigma_{urvs}$ or $C_{ijk}$, and check whether 
they still disappear in the same way as before. 

In the hyper-hyper block of the averaged scalar mass matrix, one finds that the term involving $R_{usvr}$
gives the following contribution:
\be
m^2_{u v} n^{ux} n^{vx} \supset - \hspace{1pt} 4\hspace{1pt} c^2 R_{u r v s} J^x{}^u{\!}_p J^x{}^v{\!}_q 
k_\alpha^r k_\beta^s k_\gamma^p k_\delta^q L^\alpha \bar L^\beta L^\gamma \bar L^\delta \,.
\ee
We now see that the $\Sigma_{u r v s}$ part of $R_{u r v s}$ also contributes to this contraction, because this 
now involves the full contraction $R_{u r v s} J^x{}^u{\!}_p J^x{}^v{\!}_q$ while only the completely symmetric 
part of it $R_{(u r v s} J^x{}^u{\!}_p J^x{}^v{\!}_{q)}$ is fixed by the sum rule (\ref{RiemannPropHyper}).
The $\Sigma_{urvs}$ dependence thus disappears only when $n_G = 1$, or whenever all of the 
$n_G$ sections $L^\alpha$ accidentally have the same phase. 

In the vector-vector block of the averaged scalar mass matrix, one finds that the term involving $R_{i \bar \jmath p \bar q}$ gives 
the following contribution:
\be
m^2_{i \bar \jmath} w^{ix'\!} \bar w^{\bar \jmath x'\!} \supset - \hspace{1pt} 2\hspace{1pt} c^{x'}_x \!c^{x'}_y\! R_{i \bar \jmath p \bar q} 
\bar f^{\alpha i} f^{\beta \bar \jmath} \bar f^{\gamma p} f^{\delta \bar q} 
P_\alpha^x P_\beta^y P_\gamma^z P_\delta^z \,.
\ee
We now see that the $-C_{ipr} g^{r \bar s} \bar C_{\bar \jmath \bar q \bar s}$ part of $R_{i \bar \jmath p \bar q}$
also contributes to this contraction, because the $x,y,z$ indices are contracted in a way that no longer allows for any
simplification of the result by making use of the stationarity condition (\ref{Stat2HyperVectorGen}), 
which fixes the value of $C_{ikl} \bar f^{\alpha k} \bar f^{\beta l} P_\alpha^z P_\beta^z$ and thus of 
the different contraction $R_{i \bar \jmath p \bar q} \bar f^{\alpha i} f^{\beta \bar \jmath} \bar f^{\gamma p} f^{\delta \bar q} 
P_\alpha^x P_\beta^y P_\gamma^x P_\delta^y$. Therefore the $C_{ijk}$ dependence can only be eliminated through
the stationarity condition when $n_V=1$ and $n_G = 1,2$, or whenever all of the $n_G$ triplets of Killing prepotentials 
$P^x_\alpha$ are accidentally parallel.\footnote{A similar situation also arises in ${\cal N} =4$ supergravity with vector 
multiplets, where it has been found in \cite{BR} that the sGoldstino directions pick out a different set of embedding tensor 
components to those appearing in the stationarity conditions, implying the loss of simplification of the sGoldstino mass matrix.}

In the hyper-vector block of the averaged scalar mass matrix, finally, one finds that the term involving 
$C_{ijk}$ gives the following contribution, after using the equivariance conditions:
\be
m^2_{u i} n^{ux} w^{ix'\!} \!\supset\! c\hspace{1pt} c^{x'}_y\! C_{ikl} \bar f^{i \alpha} \! \bar f^{\beta k} \! \bar f^{\gamma l} 
(2 P_{\beta}^x P_\alpha^y k^w_\gamma k_{\delta w} \!-\! P^x_{\gamma} P_\alpha^y P_{\beta}^z P^z_{\delta} 
\!+\! P^x_{\delta} P_\alpha^y P_{\beta}^z P^z_{\gamma}) L^\delta \,. \hspace{-10pt}
\ee
We see here too that the $C_{ijk}$ dependence cannot be eliminated from the contraction, because the 
stationarity condition only fixes the value of $C_{ikl} \bar f^{\alpha k} \bar f^{\beta l} P_\alpha^z P_\beta^z$. 
The $C_{ijk}$ dependence can again only be eliminated through
the stationarity condition when $n_V=1$ and $n_G = 1,2$, or whenever all of the $n_G$ triplets of Killing 
prepotentials $P^x_\alpha$ are accidentally parallel. 

We conclude that whenever $n_V \geq 2$ and $n_G \geq 2$, and no accidental simplification occurs, 
there is no way of getting rid of both of the 
$\Sigma_{urvs}$ and $C_{ijk}$ tensors controlling the curvature of the scalar manifold by averaging 
over the sGoldstino directions in the two sectors. This implies that no simple universal bound on scalar 
masses can be derived and strongly suggests that the smallest eigenvalue of the scalar mass 
matrix can be freely adjusted by tuning the values of the curvature at the stationary point under 
consideration. To see that this is indeed the case, we can consider tuning the values of 
$\Sigma_{urvs}$ and $C_{ijk}$, compatibly with the constraints imposed by the stationarity 
conditions, and then check that all the mass eigenvalues can indeed be made arbitrarily large 
relative to the gravitino mass. In this respect, we first notice that the values of the independent components 
of $\Sigma_{urvs}$ are left completely unconstrained by the stationarity conditions in the hyper sector, 
while the values of the independent components of $C_{ijk}$ are only partly constrained by the 
stationarity conditions in the vector sector unless additional peculiarities arise. 
As a result, by taking the values of $\Sigma_{urvs}$ and the unfixed values of $C_{ijk}$ 
to be large, one may achieve values of ${\cal O}(\Sigma\hspace{1pt} m^2_{3/2})$ 
for all the entries of $m^2_{uv}$ and values of ${\cal O}(|C|^2 \hspace{1pt} m^2_{3/2})$ for 
all the entries of $m^2_{i \bar \jmath}$, while the entries of $m^2_{ui}$ will only be of 
${\cal O}(C \hspace{1pt} m^2_{3/2})$. After diagonalization, one then finds
$4\hspace{1pt} n_H \! -n_V$ square mass eigenvalues of ${\cal O}(\Sigma \hspace{1pt} m^2_{3/2})$ 
and $n_V$ square mass eigenvalues of ${\cal O}(|C|^2\hspace{1pt} m^2_{3/2})$, since the level repulsion 
effect induced by the off-diagonal block gives only negligible corrections (of ${\cal O}(m^2_{3/2})$ 
on the former and of ${\cal O}(|C|^2\Sigma^{-1} \hspace{1pt} m^2_{3/2})$ on the latter), 
and all of them are then large with respect to $m^2_{3/2}$. 

We have performed various checks to verify that there is no general obstruction against 
achieving the situation described above, where arbitrary values for the scalar masses 
can be found by adjusting the curvatures of the quaternionic-K\"ahler and special-K\"ahler 
manifolds. It would be very interesting to construct an explicit family of examples 
where this can be realized concretely. For instance, one could consider 
the model with one hyper and two gauged isometries which can be described in general 
by the Calderbank-Pedersen space \cite{CP}. Unfortunately, even for this simple case it turns out to 
be algebraically complex to proceed along the same lines as \cite{CSS2}. 

Finally, we can also consider the special case with one graviphoton and one vector gauging 
two isometries (i.e. $n_G = 2$ and $n_V=1$). In this case it is possible to remove the $C_{ijk}$ 
tensor but not the $\Sigma_{urvs}$ tensor from the scalar mass matrices. This implies that the 
entries of $m^2_{i \bar \jmath}$ are fixed, whereas one can still achieve values of 
${\cal O}(\Sigma\hspace{1pt} m^2_{3/2})$ for the entries of $m^2_{uv}$. One can then see 
that $\Sigma_{urvs}$ can be tuned to make $4\hspace{1pt} n_H \! -n_V$ eigenvalues large, 
while the smallest of the remaining $n_V$ eigenvalues is bounded by $m^2_{\rm vv}$. 
However, it remains unclear whether or not the value of $m^2_{vv}$ can be made arbitrarily large 
by adjusting parameters, {\it i.e.} whether or not a bound emerges in this case.

\section{Conclusions}
\setcounter{equation}{0}

In this work, we have studied the question of whether a universal bound exists on 
the scalar masses in a supersymmetry breaking vacuum of a generic $N=2$ 
supergravity theory involving both hyper and vector multiplets. We have shown that such 
a universal bound indeed exists for any theory where at most one isometry is gauged,
and depends only on the gravitino mass $m_{3/2}$ and the cosmological constant $V$ 
at the vacuum. This result generalizes various previous analyses that were carried out for simpler
restricted classes of situations involving a minimal type and/or a minimal number of 
multiplets \cite{GRLS,MANY,CSS2}, and implies that in these theories metastable de 
Sitter vacua can exist for $V \gg m^2_{3/2}$, but not for $V \ll m^2_{3/2}$. 
We then argued that such a universal bound does not exist for theories where 
two or more isometries are gauged, and that in those theories any desired values for 
the lightest scalar square mass can, in principle, be obtained by suitably adjusting the curvature of 
the scalar manifold at the vacuum point through the parameters of the model. This implies 
that in such more general theories metastable de Sitter vacua can exist not only for 
$V \gg m^2_{3/2}$, but also for $V \ll m^2_{3/2}$. 

We believe that the result presented in this paper represents a useful guideline towards 
the search for metastable de Sitter vacua or slow-roll inflationary trajectories in supergravity 
theories emerging from string models, which often have at least some of the characteristics 
of theories with extended supersymmetry, even if they display only minimal supersymmetry. 

\vskip 20pt
\noindent
{\Large \bf Acknowledgements}
\vskip 10pt
\noindent
We would like to thank Marta Gomez-Reino and Jan Louis for useful discussions.
The research of C.~S. and P.~S. is supported by the Swiss National 
Science Foundation (SNSF) under the grant PP00P2-135164, and that of F.~C.
by the Angelo Della Riccia foundation.

\end{document}